\documentclass[11pt,a4paper,oneside]{article}
\usepackage[top=1in,bottom=1in,left=.5in,right=.5in]{geometry}
\usepackage[margin=.5in,small]{caption}

\usepackage{authblk}
\usepackage{cite}
\usepackage{amsmath,gensymb}
\usepackage{epsfig}
\usepackage{graphics}
\usepackage{amsfonts, amssymb, upgreek}
\usepackage{textcomp}
\usepackage[english]{babel}
\usepackage{blindtext}
\usepackage{color}

\providecommand{\abs}[1]{\vert #1\vert}

\newcommand{\bsigma}{\boldsymbol{\sigma}}
\newcommand{\bvarrho}{\boldsymbol{\varrho}}
\newcommand{\balpha}{\boldsymbol{\alpha}}
\newcommand{\bx}{\boldsymbol{x}}
\newcommand{\by}{\boldsymbol{y}}
\newcommand{\bu}{\boldsymbol{u}}
\newcommand{\bH}{\boldsymbol{H}}
\newcommand{\cW}{\mathcal{W}}

\newcommand{\cZa}{\mathcal{Z}_\mathrm{a}}
\newcommand{\cZs}{\mathcal{Z}_\mathrm{s}}

\newcommand{\vp}{\varrho_{_{+}}}

\newcommand{\vm}{\varrho_{_{-}}}

\newcommand{\vpm}{\varrho_{_{\pm}}}

\newcommand{\Pst}{P\!_\mathrm{st}}
\newcommand{\sgn}{\mathop{}\!\mathrm{sgn}}
\newcommand{\supp}{\mathop{}\!\mathrm{supp}}

\newcommand{\upd}{\mathop{}\!\mathrm{d}}

\newcommand{\erf}{\mathrm{erf}}

\newcommand{\Or}{\mathrm{O}}
\begin{document}

\title{Coarse grained approach for volume conserving models}

\author[1]{D. Hansmann\thanks{David.Hansmann@gmx.net}}
\author[1,2]{R. C. Buceta\thanks{rbuceta@mdp.edu.ar}}

\affil[1]{Departamento de F\'{\i}sica, FCEyN, Universidad Nacional de Mar del Plata}
\affil[2]{Instituto de Investigaciones F\'{\i}sicas de Mar del Plata, UNMdP and CONICET}
\affil[{ }]{Funes 3350, B7602AYL Mar del Plata, Argentina}

\maketitle

\abstract{
Volume conserving surface (VCS) models without deposition and evaporation, as well
as ideal molecular-beam epitaxy models, are prototypes to study the symmetries of
conserved dynamics. In this work we study two similar VCS models with conserved noise, which differ from each other by the axial symmetry of their dynamic hopping rules. We use a coarse-grained approach to analyze the models and show how to determine the coefficients of their corresponding continuous stochastic differential equation (SDE) within the same universality class. The employed method makes use of small translations in a test space which contains the stationary probability density function (SPDF). In case of the symmetric model we calculate all the coarse-grained coefficients of the related conserved Kardar-Parisi-Zhang (KPZ) equation. With respect to the symmetric model, the asymmetric model adds new terms which have to be analyzed, first of all the diffusion
term, whose coarse-grained coefficient can be determined by the same method. In contrast to other methods, the used formalism allows to calculate all coefficients of the SDE theoretically and within limits numerically. Above all, the used approach connects the coefficients of the SDE with the SPDF and hence gives them a precise physical meaning.
}



\section{Introduction}
The roughening properties of nonequilibrium surface systems with well-defined power-law behavior can be classified into universality classes by a unique set of exponents, which determine their dynamical scaling \cite{Family1988}. Discrete models and continuous equations within the same universality class share the same scaling exponents, which can be shown by numerical and analytical methods. In order to determine these exponents, many discrete models and continuous stochastic differential equations (SDE) have been studied using simulations, symmetry analyses, dynamical renormalization group theory, or numerical integration \cite{Barabasi1995}. 

Discrete models and continuous SDE within the same universality class share not only the same scaling exponents, but also the same linearities and (or) nonlinearities \cite{Lai-91}. 
In this context, models belonging the Kardar-Parisi-Zhang (KPZ) \cite{Kardar1986} or to the conserved KPZ universality class are of special importance, since for these models it is possible to determine the nonlinearities via Monte Carlo simulations using the interface tilting method \cite{Krug1989, Huse-90}.
Another method, which is not limited to the calculation of the KPZ-nonlinearities, was introduced by Vvedensky {\sl et al.} \cite{Vvedensky-93}.
They start this calculation from a discrete SDE (or discrete Langevin equation) and determine the continuous counterpart of the discrete SDE using an analytical approach. The approach is based on the regularization of the Heaviside $\Theta$-function and on the coarse grain approximation using a lattice constant which tends to zero 
\cite{Park1995,Costanza1997,Buceta2005,Muraca2004}. 
The standard procedure to regularize a function is to replace each Heaviside $\Theta$-function by a smooth function $\theta_\varepsilon$, which is continuously differentiable to any order and depends on a regularization parameter ε which has to be chosen in a way that $\theta_\varepsilon\to\Theta$ when $\varepsilon\to 0$. For details, see Appendix A. The regularization function $\theta_\varepsilon$ has to be analytic throughout its domain, but especially at zero, in order to enable a Taylor series expansion. The Taylor coefficients depend on both, the regularization prescription ({\sl i.e}. the regularization function chosen) and the regularization parameter $\varepsilon$. As pointed out by Katzav and Schwartz \cite{Katzav-04} this expansion is problematic since in the limit $\varepsilon\to 0$ the Heaviside $\Theta$-function is not analytic around zero. Consequently the ε parameter cannot be removed in the process of coarse-graining. Another weak point of the method is the difficulty of reaching conclusive results in models of higher dimensions than one. Finally, although the mathematical derivation is direct, it can generate discrepancies in interpretation of coarse coefficients of the continuous differential equation.

Recently, we introduced a different coarse-grained approach based on generalized function or distribution theory \cite{Buceta-12}. We showed, that using our approach, it is possible to calculate not only nonlinear, but also, all coefficients of the stochastic differential equation for a given discrete model of the KPZ universality class. In this work we use our formalism to show, how to determine the coarse-grained coefficients of two volume conserving surface models.

There are two well-defined groups of models and equations as distinguished by their noise, which is either non-conservative or conservative. Growth models and equations which describe deposition or evaporation process are included in the first group. In contrast, volume conserving surface (VCS) models without deposition or evaporation are included in second group. The models studied in this work are VCS models and consequently have conservative noise. Volume conserving processes are defined as physical processes which occur on the surface of a solid and preserve the total volume enclosed by the surface. These processes describe the movement of a particle from a site of the surface to another, and exclude particle deposition or evaporation. The conserved noise is assumed to be Gaussian distributed and uncorrelated. It has an expectation value of zero and the correlation is
\begin{equation}
\langle\eta(\vec{x},t)\,\eta(\vec{x}',t')\rangle=-2\,Q\,\nabla^2\delta(\vec{x}-\vec{x} ')\,\delta(t-t')\;,\label{cnoise}
\end{equation}
with $\vec{x}\in\mathbb{R}^d$ and where the conserved noise intensity $Q$ is proportional to the temperature of the system. 

The conserved KPZ (cKPZ) equation [$h=h(\vec{x},t)$, $\nu<0$, and $\lambda<0$]
\begin{equation}
\frac{\partial h}{\partial t}=\nabla^2\biggl[\nu\,\nabla^2 h+\frac{\lambda}{2}\bigl(\nabla h\bigr)^2\biggr]+\eta(\vec{x},t)\;,\label{SGG}
\end{equation}
proposed by Sun {\sl et al.}~\cite{Sun-89} describes the first continuous VCS process with conserved noise. The expression in the brackets of eq.(\ref{SGG}) we call KPZ kernel of the equation. In the first molecular-beam epitaxy (MBE) models, eq.~(\ref{SGG}) reappears with an additive constant $F$ that takes the deposition flux into account. This extended equation is called Villain-Lai-Das Sarma (VLD) equation, named by the authors of the article it was first time mentioned \cite{Villain-91,Lai-91}. In contrast to the noise term of the cKPZ equation, the noise term of the VLD equation describes non-conservative noise. It has an expectation value of zero and the correlation $\langle\eta(\vec{x},t)\,\eta(\vec{x}',t')\rangle=-2\,D\,\delta(\vec{x}-\vec{x}')\,\delta(t-t')$, where the noise intensity $D$ is proportional to the flux $F$. Although the cKPZ and VLD equations have the same KPZ kernel, the equations show distinct macroscopic properties, which have been identified and studied in various scientific works \cite{Krug-97}. 
Since both equations have the same kernel, these distinct properties can be attributed the different nature of their noise.  

A VCS model can be described by the continuity differential equation or cons\-ervation law
\begin{equation}
\frac{\partial h}{\partial t}+\nabla\cdot\mathbf{j}=\eta\;,
\end{equation}
where $\mathbf{j}=\mathbf{j}(\vec{x},t)$ is the surface diffusion current and $\eta=\eta(\vec{x},t)$ is the conserved noise. The form of the current $\mathbf{j}$ is determined by the symmetries of the system. The current cannot depend explicitly on $h$, since this would break the invariance under constant height translations. It is expected that under nonequilibrium conditions the total current $\mathbf{j}$ has two terms, {\textsl{i.e.}} $\mathbf{j}=-\nabla\mu_{\mathrm{ne}} + \mathbf{j}_{\mathrm{ne}}$. Here $\mu_\mathrm{ne}$ is the nonequilibrium chemical potential and $\mathbf{j}_\mathrm{ne}$ is the nonequilibrium current. From symmetry consideration it is expected that $\mu_\mathrm{ne}$ is a function of $\nabla^2 h$ and $\nabla h$. 

The first model studied in this work has both of these terms. 
Its dynamic rules are axial symmetric, which means, that a randomly chosen particle hops independently from the actual surface configuration with the same probability to left as it hops to the right.
Its KPZ kernel [see the eq.~(\ref{SGG})] is proportional to $\mu_\mathrm{ne}$. Furthermore it is expected that $\mathbf{j}_\mathrm{ne}$ is an odd function of $\nabla h$. 

The second model studied in this work has both types of current contribution, $\nabla\mu_\mathrm{ne}$ and $\mathbf{j}_\mathrm{ne}$.
Its dynamic rules are not axial symmetric, which means, that it depends on the surface configuration whether a randomly chosen particle hops to the left or to the right. 
In order to study their models, the majority of theoretical studies employ expansions of $\mathbf j$ in powers of $\nabla h$, its derivatives, and their combinations taking into account the symmetries of the system. The most relevant term of these expansions is determined by the long wavelength limit after renormalization \cite{Lopez-05}. 

The main goal of this work is to introduce an approach that allows to determine the coarse-grained coefficients of the continuous SDE. Additionally, we show how to use the approach to calculate coarse-grained coefficients related to two different discrete VCS models, one with symmetric hopping rules and the other with asymmetric hopping rules. The paper is organized as follows. In Section~\ref{sec:VCS} we developed some basic concepts and the theoretical formalism which is used to study the chosen models. In this context we give an overview on the conserved dynamics and conserved noise.
In Section~\ref{sec:bridge} we introduce a theoretical approach that allows to determine the coarse-grained coefficients using generalized functions or distributions theory and proper test functions. We show how to obtain these distributions and the test functions. We discuss how small changes of the test functions together with a Taylor expansion of the distribution can be used to calculate the coarse-grained observables by applying the expanded distributions to test functions.
In Sections~\ref{sec:Sym} and \ref{sec:Asym} we applying the formalism on the chosen discrete VCS models. We obtain analytical expressions for the coefficients in terms of test functions. 
In Appendix~A we show how to calculate the coarse-grained coefficients using the standard coarse-grained approach {\sl via} the regularization of $\Theta$-functions. 
In Appendix~B we evaluate the results of Section~\ref{sec:Sym} and \ref{sec:Asym} numerically using Monte Carlo simulations. 
Section~\ref{sec:concl} finishes the paper with our conclusions. 

\section{Volume conserving models\label{sec:VCS}}

By definition, volume conserving surface models keep the total number of particles enclosed by the surface constant. The movement of the particles can only increase (or decrease) the height of a surface site at the expense of decreasing (or increasing) the height of another surface site. Thus, the average velocity of the surface growth is zero and expectation value of the surface height is constant. We assume a discrete process that takes place on a square lattice with unit cell of side $a$. Further it is assumed, that a randomly chosen particle can move only one unit in a unit time. The surface configuration $\bH$ of the system is determined by set of heights $\{h_j\}$ which correspond to the columns $j=1,\dots ,N$.  The heights $h_j$ of the columns
$j$ are multiples of the unit cell size $a$, {\sl i.e.} $h_j = n_j a$ such that $n_j\in \mathbb{Z}$.

The transition rate $W(\bH,\bH')$ of a VCS model, which describes the change between two consecutive surface configurations $\bH$ and $\bH'$ in the lapse $\tau$, is 
\begin{eqnarray}
&& W(\bH,\bH')=\frac{1}{\tau}\,\sum_{k=1}^N\;[\omega_k^+\,\Delta(h'_k,h_k-a)\,\Delta(h'_{k+1},h_{k+1}+a)\nonumber\\ 
&&\hspace{3.2cm}+\,\omega_{k+1}^-\,\Delta(h'_{k+1},h_{k+1}-a)\,\Delta(h'_k,h_k+a)]\prod_{j\neq k,k+1}\Delta(h'_j,h_j)\;,\label{transition}
\end{eqnarray}
where $\omega_k^+$ and $\omega_{k+1}^-$ are the hopping rates from column $k$ to column $k+1$, and from column $k+1$ to column $k$, respectively. Here $\Delta(x,y)$ is equal to 1 if $x=y$ and equal to $0$ in all other cases. The hopping rates depend on the height differences between the chosen column and its nearest-neighbor (NN) columns. Explicitly, $\omega_j^\pm$ is function of $h_{j+1}-h_j$ and $h_{j-1}-h_j$. The first transition moment is
\begin{equation}
K_j^{(1)} =\sum_{\bH'}(h'_j-h_j)\,W(\bH,\bH')=\frac{a}{\tau}\,(\omega_{j+1}^--\omega_j^--\omega_j^++\omega_{j-1}^+)\,,
\label{1st-transition}
\end{equation}
and the second transition moment, written in an equivalent form to the expression used by ref.~\cite[eq.~6]{Jung1999}, is
\begin{eqnarray}
&& K_{ij}^{(2)} = \sum_{\bH'}(h'_i-h_i)(h'_j-h_j)\,W(\bH,\bH') \nonumber\\ 
&&\hspace{5ex}=-\frac{a^2}{2\tau}\Bigl[\bigl(\omega_i^+ +\omega_i^-+\omega_{i-1}^+ +\omega^-_{i+1}\bigl)\bigl(\delta_{i+1,j}-2\,\delta_{i,j}+\delta_{i-1,j}\bigr)\nonumber\\
&&\hspace{13ex}+\bigl(\omega_i^+ -\omega_i^- -\omega^+_{i-1}+\omega_{i+1}^-\bigr) \bigl(\delta_{i+1,j}-\delta_{i-1,j}\bigr)\Bigr]\,,\label{2nd-transition}
\end{eqnarray}
where $\delta_{i,j}$ is the Kronecker delta. Note that the first term in brackets contains the symmetric contributions and the second term in brackets contains the asymmetric contributions. If the hopping rates away from a chosen column $i$ are equal, {\sl i.e.} $\omega_i^+=\omega_i^-$, and the hopping rates towards the column $i$ are also equal {\sl i.e.} $\omega_{i-1}^+=\omega_{i+1}^-$, then eq.~(\ref{transition}) has only symmetric contributions. This means, that in this case the jumping rates and the second transition moment are invariant to the exchange of \mbox{$h_{i-n}\!\leftrightarrow h_{i+n}$} (with $n=1,2$) around to column $i$. 

Starting from the Master equation of the transition probability given by eq.~(\ref{transition}) and performing a Kramers-Moyal expansion \cite{Kramers-40,Moyal-49,VanKampen-81}, it is possible to derive an associated discrete Langevin equation
\begin{equation}
\frac{\upd h_j}{\upd t}=K_j^{(1)}+\,\eta_j(t)\,,\label{discrete-L}
\end{equation}
with $j=1,\dots, N$. The expectation value of the noise $\eta_j$ is zero and the correlation is 
\begin{equation}
\langle\eta_i(t)\eta_j(t')\rangle=K_{ij}^{(2)}\,\delta(t-t')\,.\label{discrete-C}
\end{equation} 
Jung and Kim showed \cite{Jung1999} that the continuous SDE can be achieved from its discrete counterpart [eq.~(\ref{discrete-L})] following the approach introduced by Vvedensky {\sl et al.}~\cite{Vvedensky-93}.
In Appendix A we use this approach to calculate the coarse-grained coefficients 
of the models which we discuss in the Sections~(\ref{sec:Sym}) and (\ref{sec:Asym}) in terms of the regularizing coefficients. 

\section{A bridge from discrete to continuous processes of the same\\ universality class\label{sec:bridge}}

According to the previous section, the particle hopping rate $\omega_j^\pm$ from the column $j\to j\pm 1$ is a function of the integer set $\{\sigma_{j\pm}=(h_{j\pm 1}-h_j)/a\}$. Thus, the transition moments $K_j^{(1)}$ and $K_{ij}^{(2)}$ [eqs.~(\ref{1st-transition}) and (\ref{2nd-transition}), respectively] depend on the hopping rates and the hopping rates are functions of the height differences between the NN columns and the selected column. The identity $\omega_j^+=\omega_j^-$ is a fundamental property of the studied system which is indispensable to extract all the coarse-grained properties derived from the transition moments. 
Using the transition rates $\omega_j^\pm$ we define a generalized function $\mathcal{W}(\bvarrho)$ with $\bvarrho\in\mathbb{R}^2$. We assume, that this generalized function matches with the hopping rates, {\sl i.e.} $\mathcal{W}(\bvarrho=\bsigma_j)=\omega_j^\pm$ with $\bsigma_j\in\mathbb{Z}^2$. The generalized function $\mathcal{W}$ can be decomposed in a symmetric $\mathcal{W}_\text{s}$ and an antisymmetric $\mathcal{W}_\text{a}$ term in order to calculate coarse-grained observables.

\begin{figure}[t!]
\centerline{\includegraphics[width=0.43\linewidth]{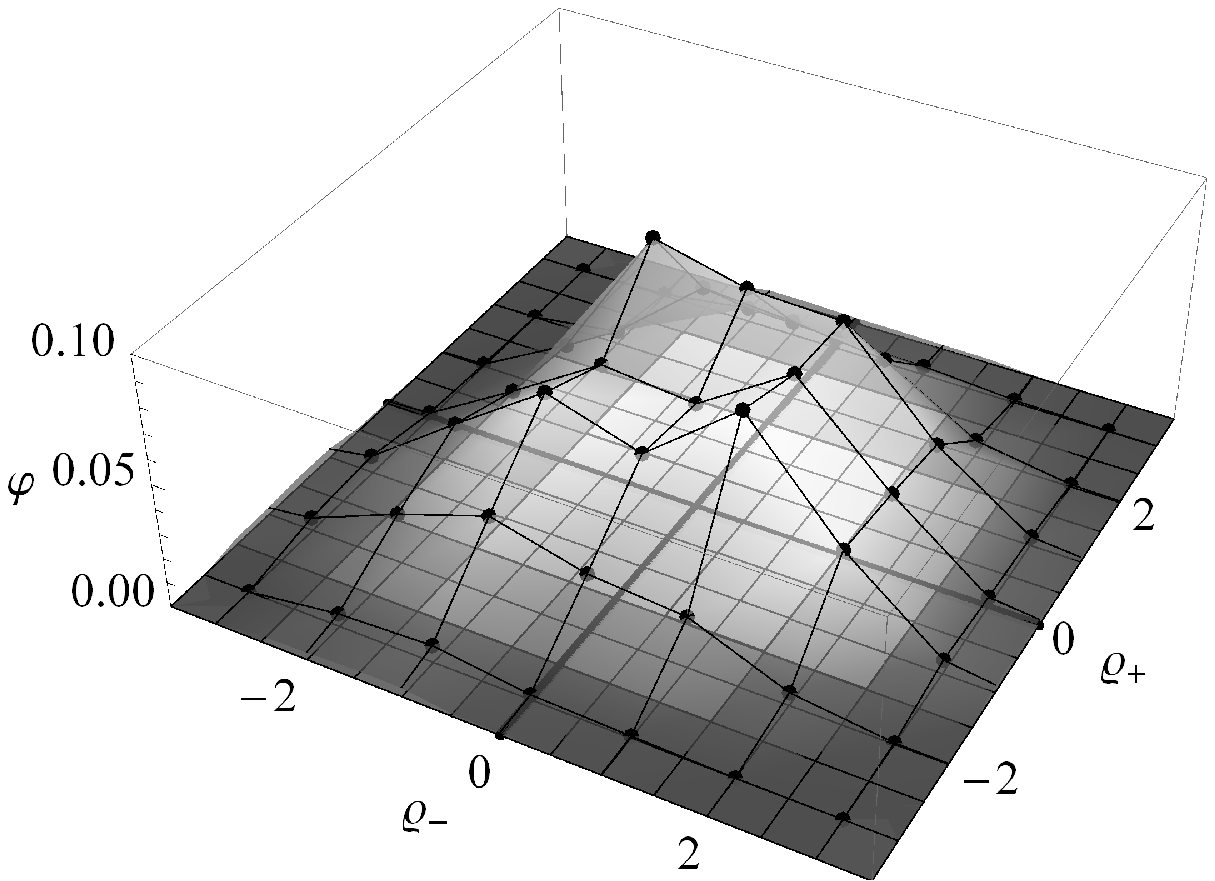}
\includegraphics[width=0.43\linewidth]{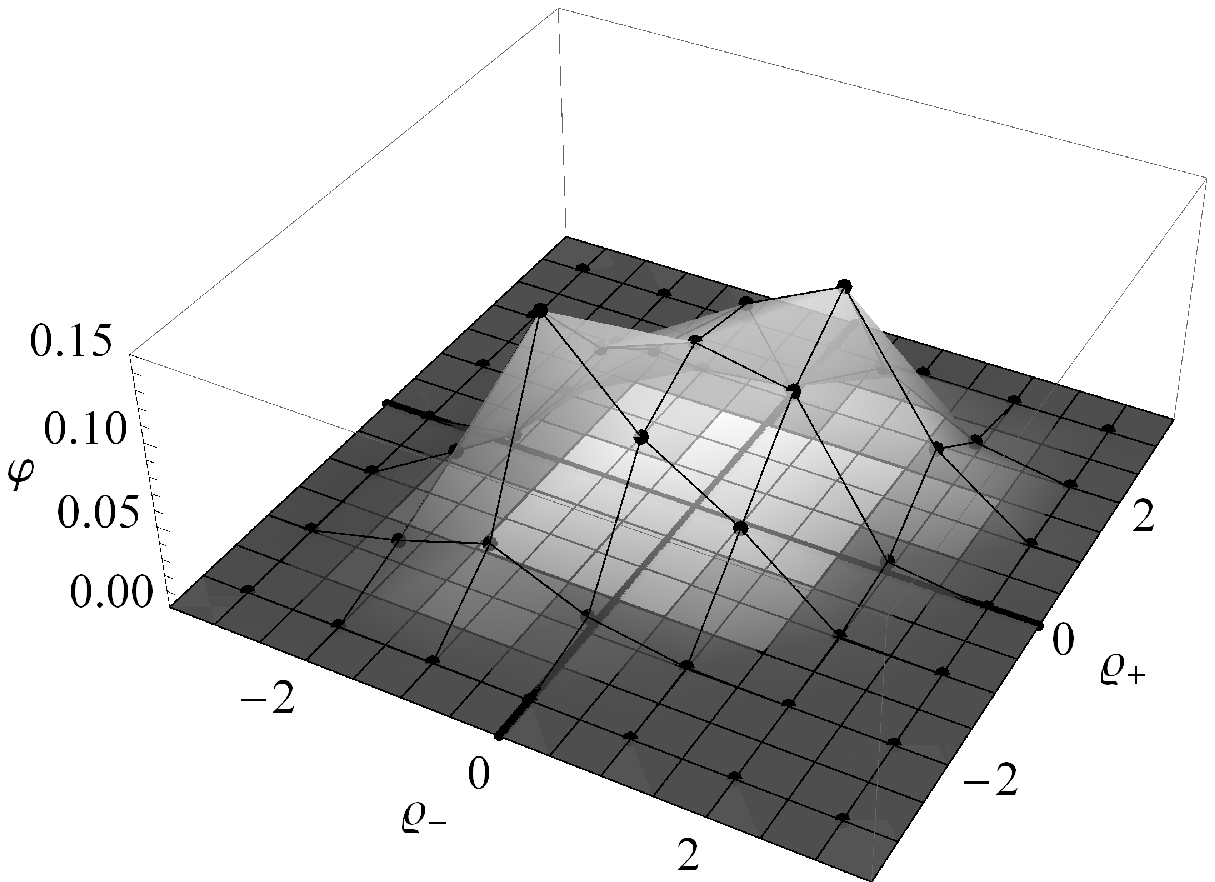}}
\caption{
The figure shows the test space. Left: symmetric hopping model. Right: asymmetric hopping model. In the test space one sees a test function
$\varphi(\bvarrho_{+},\bvarrho_{-})$ (semi-transparent) and the discrete probability density function $\Pst(\sigma_{+},\sigma_{-})$ (black bullets). One sees, that the test function matches with the discrete probability density function for $\bvarrho=\sigma$. The test function $\varphi$ is smooth though not visually fit.}\label{fig1}
\end{figure}

In order to calculate the statistical observables of discrete processes it is necessary to define test functions $\varphi$ on which the generalized functions of $\bvarrho$ can be applied.  These test functions  $\varphi$ have to have different properties for restricted and unrestricted processes. Restricted processes require that $\varphi\in\mathcal{D}(\mathcal{R}^2_\varepsilon)$, where $\mathcal{D}$ is the test space of $\mathrm{C}^\infty$-functions with compact support and $\mathcal{R}^2_\varepsilon=\varepsilon\mathrm{-neighborhood}(\mathcal{R}^2)$.
Unrestricted processes require that $\varphi\in\mathcal{S}(\mathbb R^2)$, where $\mathcal{S}$ is the test space of $\mathrm{C}^\infty$-functions that decay and have derivatives of all orders that vanish faster than any power of $\varrho_\alpha^{-1}$ ($\alpha=1,2$). In the present work we define the test functions on the base of the surface configuration. For each column $j$ we have two random integer variables that define a surface configuration vector $\bsigma_j=(\sigma_{j-},\sigma_{j+})\in\mathbb{Z}^2$. The chance to find a column $j$ with the surface configuration $\bsigma_j$ at the time $t$ is given by the time dependent probability density function $P(\bsigma_j,t)$. One can show via Monte Carlo simulations, that the time dependent probability density function converges rather fast. Hence, for $t\gg 1$ one can interpret $P(\bsigma_j,t)$ as a time independent, stationary probability density function $\Pst(\bsigma_j)$ (SPDF). The set of surface configurations in the steady state is $\mathcal{Z}^2 =\{\bsigma_j\in\mathbb{Z}^2 / \Pst(\bsigma_j)\neq 0,\,\forall j=1,\dots, N\}\,$. In the present work the steady state configuration space $\mathcal{Z}^2$ is a 2-dimensional lattice within a compact set $\mathcal{R}^2\subset(\mathbb{R}^2\setminus\mathbb{Z}^2)\cup\mathcal{Z}^2$. We define the test function $\varphi(\bvarrho)$ as a real-valued function, that matches with the discrete SPDF, {\sl i.e.}  $\varphi(\bvarrho=\bsigma_j)=\Pst(\bsigma_j)$ with $\bsigma_j\in\mathbb{Z}^2$. Fig.~\ref{fig1} shows the test functions of both the unrestricted VCS models considered in this work.

The application of a distribution $f\in\mathcal{D}'$ (dual space of $\mathcal{D}$) to test function $\varphi\in \mathcal{D}$ is defined by
\begin{equation}
\langle f\,,\varphi\rangle=\int_{\mathbb{R}^2}f(\bvarrho)\,\varphi(\bvarrho)\;\mathrm{dv}_{\!\varrho}\;.\label{distr_f}
\end{equation}
We used here the notation on distributions that was introduced by Schwartz \cite{Schwartz-66}. 
Note that $\langle f,\varphi\rangle$ is the ``expectation'' value of $f$ using the test function $\varphi$ as real-valued analytic ``representation'' of the SPDF $\Pst$. Thus, the test function is normed, {\sl i.e.} $\langle 1,\varphi\rangle=1\,$.
The translation $T_{\balpha}$ of a distribution $f$, denoted $T_{\balpha}f$ or simply $f_{\balpha}$, extends the definition given by eq.~(\ref{distr_f})
\begin{equation}
\langle T_{\balpha} f\,,\varphi\rangle=\langle f\,, T_{-\balpha}\,\varphi\rangle\,,
\end{equation}
where the translation operator is defined by $T_{\bx}:\by\mapsto\by -\bx$ if $\by\,,\bx\in\mathbb{R}^2$ \cite{Colombeau1984}. As mentioned above, we assume that the test function $\varphi$ takes ​​fixed values ​​in the discrete lattice $\mathcal{Z}^2$ given by the SPDF $\Pst$, {\sl i.e.} 
$\langle \delta_{\bsigma}\,,\varphi\rangle=\varphi(\bsigma)=\Pst(\bsigma)\,$
for all $\bsigma\in\mathcal{Z}^2\,$, where $\delta_{\bsigma}=T_{\bsigma}\delta$ and $\delta$ is the Dirac distribution.

Applying a translation $T_{\bu}$ to a point $\bvarrho\in\supp(\varphi)$, the test function transforms as $\varphi\rightarrow T_{\bu}\varphi$\, if\,  $(\bvarrho-\bu)\in\supp(\varphi)$. Furthermore $w=\langle\cW,\varphi\rangle$, the expectation value of $\cW$, changes as $w(0)\rightarrow w(\bu)$ with
\begin{equation}
w(\bu)=\langle\cW\,,T_{\bu}\varphi\rangle=\int_{\mathbb{R}^2}\cW(\bvarrho)\;\varphi(\bvarrho-\bu)\;\mathrm{dv}_{\!\varrho}\;.
\label{eq:w-u}
\end{equation}
For small translations, the Taylor expansion of $\varphi(\bvarrho-\bu)$ around $\bu =\boldsymbol 0\,$ is
\begin{equation}
\varphi(\bvarrho-\bu)=\varphi(\bvarrho)-u_\alpha\;\partial_\alpha\varphi\bigr\rfloor_{{\bu} =\mathbf{0}}+\tfrac{1}{2}\;u_\alpha u_\beta\;\partial^2_{\alpha\beta}\varphi\bigr\rfloor_{{\bu} =\mathbf{0}}+\Or(3)\;.\label{Tvarphi}
\end{equation}
Here repeated subscripts imply sums, $u_\alpha$ is the $\alpha$-th component of $\bu$, $\partial_\alpha\,\dot=\,\partial/\partial\varrho_\alpha$, and $\partial^2_{\alpha\beta}\,\dot=\,\partial^2/(\partial\varrho_\alpha\partial\varrho_\beta)$, with $\alpha\,,\beta=1,2$. 
Since the test function $\varphi$ is known only at points of the lattice $\mathcal{Z}^2$, its derivatives can not be calculated explicitly. In contrast, the distribution $\cW$ is derivable in all points. Since the test function has either compact support or decreases rapidly one can take advantage of the following identity 
\begin{equation}
\bigl\langle \cW\,,\partial^{\rm{n}}_{\alpha\beta\cdots\omega}\varphi\bigr\rangle=(-1)^{\rm{n}}\bigl\langle \partial^{\rm{n}}_{\alpha\beta\cdots\omega} \cW\,,\varphi\bigr\rangle\;.\label{repl}
\end{equation}
Using eqs.~(\ref{eq:w-u}--\ref{repl}), the observable is transformed according to
\begin{equation}
w(\bu)=\Bigl\langle\cW,\varphi\Bigr\rangle +\Bigl\langle\partial_\alpha\cW,\varphi\Bigr\rangle\,u_\alpha
+\tfrac{1}{2}\Bigl\langle\partial^2_{\alpha\beta}\cW,\varphi\Bigr\rangle\,u_\alpha u_\beta+\Or(3)\,.\label{distr-exp0}
\end{equation}

In a previous work \cite{Buceta-12} we used the mentioned translations to determine the coarse-grained coefficients for a restricted and an unrestricted discrete KPZ-type model: the restricted solid-on-solid model and the ballistic deposition model, respectively. We derive their coefficients from the transformed average velocity of the interface $v(\bu)=\langle K^{(1)},T_{\bu}\varphi\rangle$ where the first transition moment $K^{(1)}$ is the drift of corresponding Langevin equation. The VCS models studied in the following sections have no restrictions, although the theory allows the treatment of models with restrictions, as shown in previous studies.
In contrast to this former work, the presently studied models allow to calculate analytically another observable quantity which is the non-conserved noise intensity. This is possible by intervening in the equivalent continuous process, in this case via the application of small changes to $D=\langle K^{(2)},\varphi\rangle$, where $K^{(2)}$ is the second transition moment corresponding to discrete process and $D$ is the noise intensity corresponding to equivalent continuous process. 

\section{Symmetric hopping model (Krug Model)\label{sec:Sym}}

At first we study a model with symmetric hopping rate suggested by Krug \cite{Krug-97} and studied by Jung and Kim \cite{Jung1999}. The hopping rules are as follows:

Choose column $j$ randomly, if the surface configuration satisfies the condition $h_{j\pm 1}>h_j$, the surface remains unchanged and a new column is chosen. If the condition $h_{j\pm 1}>h_j$ is not fulfilled, the particle on the surface of the column $j$ moves with an equal probability either to the left or to the right neighboring column. From the rules one can see, that a particle is mobile even if the surface is totally flat. This is necessary in order to generate a nonequilibrium contribution  since there is no particle deposition. Introducing the hopping rates $w_k^\pm$, from the chosen column $k$ to right or left NN column, the first moment of the transition probability between two surface configurations is given by eq.~(\ref{1st-transition}) where the hopping rates from column $j$ are
\begin{equation}
w_j^\pm=\tfrac{1}{2}\,\Theta(h_j-h_{j-1})\,\Theta(h_j-h_{j+1})\;.\label{hopping}
\end{equation}
The first moment of transition rate is
\begin{equation}
K^{(1)}_j=a^{-2}\Delta^{\!(2)}\cZs(\vm,\vp)\Bigr\rfloor_{\vpm=(h_{j\pm 1}-h_j)/a}\,,\label{K1s}
\end{equation}
where $\Delta^{\!(2)}$ is the second variation [{\sl i.e.} \mbox{$\Delta^{\!(2)}F_j=F_{j+1}-2\,F_j+F_{j-1}$}], with the property that $a^{-2}\Delta^{\!(2)}\to\partial^2/\partial x^2$ when $a\to 0$. Here, the kernel $\cZs\!:\mathbb{R}^2\to\mathbb{R}$ is the generalized function defined by
\begin{equation}
\cZs(\vm,\vp)=\frac{\alpha}{2}\,\Theta(-\vm)\,\Theta(-\vp)\;.\label{cZs}
\end{equation}
with $\alpha=a^3/\tau$. Notice that $\cZs$ is antisymmetric, {\sl i.e.} $\cZs(x,y)=\cZs(y,x)$.
The real-valued generalized function $\cZs$ can be achieved from its discrete counterpart $w_j^\pm$ by regularization techniques. Via the coarse approximation a stochastic differential equation for  $h=h(x,t)$
\begin{equation}
\frac{\partial h}{\partial t}=\nabla^2 Z_\mathrm{s}+\eta(x,t)\;,\label{cKPZ}
\end{equation}
can be obtained, where the KPZ kernel is
\begin{equation}
Z_\mathrm{s}=\nu\,\nabla^2 h +\frac{\lambda}{2}\,\abs{\nabla h}^2\;.\label{KPZ-kernel}
\end{equation}
Here and below, we use the following notation for the partial derivative $\nabla\doteq \partial/\partial x$. Equation~\ref{cKPZ} is a conserved Kardar-Parisi-Zhang equation with conservative noise \cite{Sun-89} whose expectation value is zero and with the correlation function given by eq.~(\ref{cnoise}).
In the Appendix A we find this equation using coarse-grained techniques {\sl via} regularization by an alternative way to reference~\cite{Jung1999}. 
The coefficients $\nu$ and $\lambda$ of KPZ kernel in terms of the regularizing constants are given in eqs.~(\ref{nu}) and (\ref{lamda}), respectively. 

Instead of using coarse-grained techniques {\sl via} regularization, the coefficients $\nu$ and $\lambda$ can be obtained by applying the distribution $\cZs$ [eq.~(\ref{cZs})] on translated test functions $T_{\bu}\varphi$. 
We introduce the transformations $\vpm\rightarrow\vpm+u_{_{\pm}}\,$ and we assume that gradient and Laplacian transform with
$u_{_{-1}}\!-u_{_{+1}}=2\,s$\quad and\quad $u_{_{-1}}\!+u_{_{+1}}=2\,r\,$, respectively.
Considering a translation $\bu=(r+s,r-s)$ and taking into account that $\varphi$ is symmetric under exchange $\vm\leftrightarrow\vp$ one calculates
\begin{equation}
z_\mathrm{s}(r,s) = \bigl\langle\cZs\,,T_{\bu}\varphi\bigr\rangle = \omega_0+\nu\,r+\tfrac{1}{2}\,\lambda\,s^2+\Or(r^2\!,s^3)\;,
\end{equation}
where
\begin{eqnarray}
\omega_0\! &=& \langle\cZs,\varphi\rangle\;,\nonumber\\
\nu &=& 2\,\langle\partial_\ell\cZs\,,\varphi\rangle\;,\\
\lambda &=& 2\,\langle(\partial^2_{\ell\ell}-\partial^2_{\ell k})\cZs\,,\varphi\rangle\;,\nonumber
\end{eqnarray}
with $\ell=k=\pm 1$ and $\ell\neq k$, where repeated subscripts do not imply sums. Here $\varphi(x,y)$ is the (symmetric) test function. For discrete values $x,y\in\mathbb{Z}$ the test function $\varphi(x,y)$ takes the same value as the probability density function $\Pst(x,y)$. Taking in to account that
\begin{eqnarray*}
\partial_\ell\cZs &=& -\tfrac{1}{2}\,\alpha\,\delta(\varrho_\ell)\,\Theta(-\varrho_k)\;,\nonumber\\
\partial^2_{\ell k}\cZs &=& \tfrac{1}{2}\,\alpha\,\delta(\varrho_\ell)\,\delta(\varrho_k)\;,\\
\partial^2_{\ell\ell}\cZs &=& -\tfrac{1}{2}\,\alpha\,\delta'(\varrho_\ell)\,\Theta(-\varrho_k)\;,\nonumber
\end{eqnarray*}
we obtain
\begin{eqnarray}
\omega_0\!&=&\frac{\alpha}{2}\int\!\int_{-\infty}^0\,\varphi(x,y)\upd x\upd y\;,\label{xomega0}\\
\nu &=& -\alpha\int_{-\infty}^0\varphi(0,y)\;\upd y\;,\label{xnu}\\
\lambda &=& -\alpha\biggl[\varphi(0,0) -\int_{-\infty}^0\partial_x\varphi\Bigr\rfloor_{(0,y)}\,\upd y\biggr]\;.\label{xlambda}
\end{eqnarray}
We can easily show that the noise intensity of correlation function [eq.~(\ref{cnoise})] is 
$Q=a^2\,z_\mathrm{s}(\bu)=a^2\,\omega_0+\Or(1)$ for all translation $\bu$ with \mbox{$\abs{\bu}\ll 1$}. When $a\to 0$ the noise intensity is
\begin{equation}
Q=a^2\,\omega_0\;.\label{xQ}
\end{equation}
Please note, that using our approach we have obtained the three coarse-grained coefficients which characterize the conserved KPZ equation. The algebraic signs of these results are in agreement with the algebraic signs obtained using the coarse-grained approach {\sl via} regularization \cite{Jung1999}.

\section{Asymmetric hopping model\label{sec:Asym}}

In this section we study a model with asymmetric hopping rate introduced by Jung and Kim \cite{Jung1999}. In contrast to the previous studied model where a chosen particle moves, independently from the local surface configuration with the same probability to the left or to the right side, here a chosen particle moves to the NN column which has the lower height. If the heights of the NN columns are both lower than the height of the selected column, the particle hops randomly to one of them. Thus the symmetry of the Krug model is broken.
The hopping rates for this model are
\begin{equation}
\omega_k^\pm=\tfrac{1}{2}\,\Theta(h_k-h_{k-1})\,\Theta(h_k-h_{k+1})\,\bigl[1-\sgn(h_{k\pm 1}-h_{k\mp 1})\bigr]\;,
\label{hopping-Jung}
\end{equation}
where $\sgn(z)=\Theta(z)-\Theta(-z)$ is the sign function. The first and second term are the symmetric and antisymmetric contributions to hopping rate $\omega_j^\pm$ under the exchange \mbox{$h_{j-1}\leftrightarrow h_{j+1}$} respectively. In contrast to the hopping rate $\omega_j^\pm$, the hopping rate $\omega_{j\mp 1}^\pm$ is asymmetric under the exchange \mbox{$h_{j-n}\leftrightarrow h_{j+n}$ ($n=1,2$)}.
The first moment of the transition rate is
\begin{equation}
K^{(1)}_j=\Bigl[\frac{1}{a^2}\Delta^{\!(2)}\cZs(\vm,\vp)+\frac{1}{2a}\Delta^{\!(1)}_2\cZa(\vm,\vp)\Bigr]\biggr\rfloor_{\vpm=(h_{j\pm 1}-h_j)/a}\,,\label{K1a}
\end{equation}
where $\cZs$ is the symmetric contribution known from the Krug model given by eq.~(\ref{cZs}) and $\Delta^{\!(1)}_2$ is the first variation between the NN columns [{\sl i.e.} $\Delta^{\!(1)}_2 F_j=F_{j+1}-F_{j-1}$] with the property $(2\,a)^{-1}\Delta^{\!(1)}_2\to\partial/\partial x$ when $a\to 0$. Just like in the case of the Krug model, the kernel $\cZa\!:\mathbb{R}^2\to\mathbb{R}$ is interpreted as a generalized function defined by
\begin{equation}
\cZa(\vm,\vp)=-\beta\,\Theta(-\vp)\,\Theta(-\vm)\,\sgn(\vp-\vm)\;.\label{cZa}
\end{equation}
with $\beta=a^2/\tau$. Notice that $\cZa$ is antisymmetric, {\sl i.e.} $\cZa(x,y)=-\cZa(y,x)$.

By regularization techniques and the coarse approximation a stochastic differential equation can be obtained. In this case [$h=h(x,t)$] is
\begin{equation}
\frac{\partial h}{\partial t}=\nabla^2 Z_\mathrm{s}+\nabla\cdot Z_\mathrm{a} +\eta(x,t)\;,\label{cKPZm}
\end{equation}
where the antisymmetric kernel is
\begin{equation}
Z_\mathrm{a}=\nu_0\,\nabla h +\gamma\,\nabla h\,|\nabla h|^2\;,\label{antisym}
\end{equation}
and $Z_\mathrm{s}$ is the KPZ kernel eq.~(\ref{KPZ-kernel}) with coefficients given by eqs.~(\ref{xomega0})--(\ref{xlambda}). 

We perform a translation $\bu=(s,-s)$ on the surface configuration space with $s\ll 1$, taking into account that $\varphi$ is symmetric in its variables
\begin{equation}
z_\mathrm{a}(s) = \langle\cZa\,,T_{\bu}\varphi\rangle = \nu_0\,s+\gamma\,s^3+\Or(5)\;,\label{za}
\end{equation} 
where $z_\mathrm{a}(s)=$ and
\begin{eqnarray}
\nu_0 &=& \bigl\langle(\partial_x-\partial_y)\cZa\,,\varphi\bigr\rangle\;,\nonumber\\
\gamma &=& \bigl\langle(\partial_x-\partial_y)^{\!(3)}\!\cZa\,,\varphi\bigr\rangle\;.
\end{eqnarray}
Eq.~(\ref{za}) contains only odd powers in $s$ since $\cZa$ is antisymmetric and contributes only to their odd order derivatives which are symmetrical. Below we show only the first term, although the calculations can be extended to higher order. Taking into account that
\begin{equation*}
(\partial_x-\partial_y)\cZa(x,y)=\bigl[\delta(x)\Theta(-y)-\delta(y)\Theta(-x)\bigr]\sgn(x-y)-4\,\Theta(-x)\Theta(-y)\,\delta(x-y)\,,
\end{equation*}
we obtain
\begin{equation}
\nu_0=2\,\beta\int_{-\infty}^{0}\bigl[\varphi(0,x)-2\,\varphi(x,x)\bigr]\,\upd x\;.\label{xnu0}
\end{equation}

\section{Conclusions\label{sec:concl}}

In this work we describe a new approach to calculate coefficients of continuous SDE, which is based on well-established concepts of the theory of generalized functions.
We apply this approach and show how to calculate the coarse-grained coefficients of continuous SDE from two discrete VCS models. For reasons of clarity and comprehensibility both models are (1+1)-dimensional particle models, which differ from each other by the axial symmetry of their dynamic hopping rules. 
For the calculation we take advantage of symmetries in the transition moments of the models, which describe the particle hopping rates from one randomly selected column to its next neighbor columns.
The hopping rates depend only on the height difference between the selected and the next neighbor columns. 
We extend these discrete hopping rates (dynamic evolution rules) to the continuum by interpreting them as generalized functions or distributions. 
Finally we apply the generalized functions to test functions, where the test functions are calculated on the base of the SPDF of the surface configurations.

The introduced approach has the advantage to unveil the symmetry properties implied by a particular model and to determine the coarse-grained observables directly from the evolution rules. Although it is impossible to use the approach to determine directly the SDE, it can be used to calculate the coefficients of this equation employing a test function which is based on the SPDF. In this way the approach gives a very precise physical meaning to these coefficients, by relating them to the SPDF.

In contrast to using the classical coarse grained approach the obtained coefficients of the SDE are dependent on the regularizing coefficients, thus are dependent on the regularizing function and can not be univocally, numerically determined. 
Using our approach we give a precise meaning to these regularizing coefficients and determine them univocally relating them to the coarse-grained coefficients of both formalisms. 

We believe that this formalism can be extended to conserved dynamic models with non-conserved noise, such as Wolf-Villain \cite{Wolf-90} and Das Sarma-Tamborenea  \cite{Sarma-91} models of MBE. But, the crossover between universality classes, the change in the continuous SDE and therefore in their coefficients, suggest that the notion of configuration space introduced here must be revised for these models.
The DT model corresponds to the VLD equation, as can be proved by the method of
coarse-graining via regularization \cite{Huang-96}. The VLD equation contains the following terms: cKPZ, the flux F , and non-conserved noise. The Krug model corresponds to the SDE that has only cKPZ term and conserved noise. It is an open issue to explain
the connection through $F\to 0$, changing the noise from non-conserved to conserved.

\section*{Appendix A\label{apA}}

In this appendix we show how to obtain the continuous SDE starting from the discrete volume conserving models and using the coarse-grained approximation {\sl via} regularization of Heaviside-functions. We use a slightly different procedure than other authors, but obtain the same coefficients and equations. The expressions obtained here
are related with the expressions obtained in Sections~\ref{sec:Sym} and \ref{sec:Asym} by the coarse-grained approximation {\sl via} generalized functions introduced in Section~\ref{sec:bridge}.

 Using the regularization procedure, the Heaviside function $\Theta(x)$ can be replaced by a smooth real-valued function $\theta_\varepsilon(x)$, which satisfies $\theta_\varepsilon(n)\to\Theta(n)$ when $\varepsilon\to 0^+$ for all $n\in\mathbb{Z}$. There have been several proposals to represent $\Theta$ for a shifted analytic function, including the following 
\begin{equation*}
\theta_\varepsilon(x)=\frac{1}{2}\int_{-\infty}^x\;\Bigl[\erf\Bigl(\frac{s+1}{\varepsilon}\Bigr)-\erf\Bigl(\frac{s}{\varepsilon}\Bigr)\Bigr]\upd s\;,
\end{equation*}
with $\varepsilon >0$, introduced in refs.~\cite{Haselwandter-06,Haselwandter-07} that result well. 
The kernels $\cZs$ and $\cZa$ [eqs.~(\ref{cZs}) and (\ref{cZa}), respectively] can be regularized using the $\varepsilon$-theta function $\theta_\varepsilon$, {\sl i.e.}
\begin{eqnarray}
&&\cZs(\vm,\vp)=\tfrac{1}{2}\,\alpha\,\theta_\varepsilon(-\vm)\,\theta_\varepsilon(-\vp)\nonumber\\
&&\cZa(\vm,\vp)=-\beta\,\theta_\varepsilon(-\vm)\,\theta_\varepsilon(-\vp)\,[\theta_\varepsilon(\vp-\vm)-\theta_\varepsilon(\vm-\vp)]\;,\nonumber
\end{eqnarray}
with $\alpha=a^3/\tau$ and $\beta=a^2/\tau$. Expanding the $\varepsilon$-theta in Taylor series around $x=0$
\begin{equation}
\theta_\varepsilon(x)=\sum_{k=0}\,A_k^{(\varepsilon)}x^k\;,\label{e-theta}
\end{equation}
gives the following expansions (superscripts are omitted hereafter)
\begin{eqnarray}
&&\cZs(\vm,\vp)=\frac{1}{2}\,\alpha\biggl[A_0^2-A_0 A_1 \bigl(\vm\!+\vp\bigr)
\nonumber\label{cetas}\\&&\hspace{2.5cm}
+\frac{(2A_0 A_2-A_1^2)}{2(\gamma+1)}\bigl(\vm^2-2\gamma\vm\vp+\vp^2\bigr)+\Or(3)\biggr]\\
&&\cZa(\vm,\vp)=-\beta\Bigl[2\,A_0^2 A_1\bigl(\vp-\vm\bigr)+\Or(3)\Bigr]\nonumber\;,
\end{eqnarray}
where 
\begin{equation}
\gamma=-\frac{A_1^2}{2 A_0 A_2}\,.\label{gamma}
\end{equation}
Evaluating eqs.~(\ref{cetas})
\begin{eqnarray}
&&\cZs(\vm,\vp)\Bigr\rfloor_{\vpm=(h_{j\pm 1}-h_j)/a}=\omega_0+\nu\,L_i^{(1)}+\frac{\lambda}{2}\,N_i^{(\gamma)}+\Or(3)\nonumber\\
&&\cZa(\vm,\vp)\Bigr\rfloor_{\vpm=(h_{j\pm 1}-h_j)/a}=\nu_0\,L_i^{(2)}+\Or(3)\;,
\end{eqnarray}
where the constants are
\begin{eqnarray}
\omega_0&=&\frac{a^3}{2\,\tau}A_0^2\label{omega}\;,\\
\nu&=&-\frac{a^4}{2\,\tau}A_0 A_1\label{nu}\;,\\
\lambda&=&\frac{a^3}{2\,\tau}(2A_0 A_2-A_1^2)\;,\label{lamda}\\
\nu_0&=&-\frac{4 a^2}{\tau} A_0^2 A_1\;,\label{nu0}
\end{eqnarray}
and the linear and quadratic terms are
\begin{eqnarray}
L_j^{(1)}&=&\frac{h_{j+1}-2 h_j+h_{j-1}}{a^2}\;,\nonumber\\
L_j^{(2)}&=&\frac{h_{j+1}-h_{j-1}}{2\,a}\;,\label{discrete}\\
N_j^{(\gamma)} \!&=&\! \frac{(h_{j+1}-h_j)^2-2\,\gamma\, (h_{j+1}-h_j)(h_{j-1}-h_j)+(h_{j-1}-h_j)^2}{2 \,a^2(\gamma+1)}\;,\nonumber
\end{eqnarray}
with $0\le\gamma\le 1$ the discretization parameter of the discretized nonlinear term \cite{Buceta2005}. The usual choice $\gamma=1$ is called standard or post-point discretization. It depends only on the height of the NN columns and thus the error of approximating $(\nabla h)^2$ is minimized. In contrast, the choice $\gamma=0$, called antistandard or prepoint discretization. It corresponds to the arithmetic mean of the squared slopes around the interface sites. In this work, $\gamma$ depends on the coefficients of regularization through eq.~(\ref{gamma}), as $A_0>0$ then $A_2<0$, that coincides with the result obtained in ref.~\cite{Jung1999}. In addition we found if $A_1>0$, then $\nu<0$, $\lambda<0$, and $\nu_0<0$.
Expanding eqs.~(\ref{discrete}) around $x=ja$, the discretized terms and their limits when $a\to 0$ are
\begin{eqnarray}
L_j^{(1)}&=&\nabla^2 h+\frac{1}{12}\nabla^4 h\;a^2+\Or(4)
\quad\longrightarrow\quad\nabla^2 h\;,\nonumber\\
L_j^{(2)}&=&\nabla h+\frac{1}{16}\nabla^3 h\;a^2+\Or(4)
\quad\longrightarrow\quad\nabla h\;,\label{continuous}\\
N_j^{(\gamma)} &=&(\nabla h)^2+\frac{1}{4}\biggl(\frac{1-\gamma}{1+\gamma}\biggr)(\nabla^2 h)^2\;a^2+\Or(4)
\quad\longrightarrow\quad(\nabla h)^2\;.\nonumber
\end{eqnarray}
Notice that the limit of $N_j^{(\gamma)}$ does not depend on the discretization parameter $\gamma$, as shown in ref.~\cite{Buceta2005}. Take into account that $\frac{a^3}{\tau}\cZs\to Z_\mathrm{s}$ and $\frac{a^2}{\tau}\cZa\to Z_\mathrm{a}$ when $a\to 0$. Applying these limits to eqs.~(\ref{K1s}) and (\ref{K1a}) we obtain eqs.~(\ref{cKPZ}) and (\ref{cKPZm}), respectively. A dimensional analysis of these last continuous differential equations shows that the their coefficients given by eqs.~(\ref{omega})--(\ref{nu0}) have the correct dimensions. In order to obtain the continuous limit of eq.~(\ref{discrete-C}) we take $a\to 0$. The following limits are calculated for eq.(\ref{2nd-transition}):
\begin{eqnarray*}
a^{-3}\bigl(\delta_{i+1,j}-2\,\delta_{i,j}+\delta_{i-1,j}\bigr)&\!\longrightarrow\! &\nabla^2 \delta(x-x')\,,\\
a^5\,\tau^{-1}\bigl(\omega_i^+ +\omega_i^-+\omega_{i-1}^+ +\omega^-_{i+1}\bigr)&\!\longrightarrow\! & 4\,Q\;.
\end{eqnarray*}
Using these limits we conclude, that \mbox{$K_{i,j}^{(2)}\to K^{(2)}(x-x')=-2\,Q\,\nabla^2\delta(x-x')$}, and the continuous noise correlation function is given by eq.~(\ref{cnoise}). Finally, the noise intensity is
\begin{equation}
Q=\frac{a^5}{2\,\tau}\,A_0^2\;.
\end{equation}

\section*{Appendix B\label{apB}}

In this appendix we give the results of Monte Carlo simulations performed for the studied volume conserving models. We calculate the coarse-grained coefficients {\sl via} theory of generalized functions. Both the studied models evolve over time unit $\tau=1 $ in a square lattice with unit cell size $a = 1$, with a total system size $N$ and use periodic boundary conditions. On the discrete lattice we calculate the stationary joint probability distribution $P_\mathrm{st}(\sigma_{_-},\sigma_{_+})$ of the surface configurations with \mbox{$\sigma_{_\pm}=h_{j\pm1}-h_j$} in order to evaluate the coarse-grained coefficients (see Table~\ref{t1}). The integrals of the test function $\varphi$ or their derivatives are evaluated by conventional finite difference methods, based on $P_\mathrm{st}$ data.
\begin{table}[h!]
\begin{center}
\begin{tabular}{| l | c | c | c | c | c |}
\hline
Model & $\omega_0$ & $\nu$ & $\lambda$ & $\nu_0$ & $Q$\\ \hline\hline
SHM   & 0.180 &-0.116 & -0.034 & - & 0.180 \\ \hline
AHM   & 0.250 &-0.193 & -0.158 & -0.615 & 0.250 \\ \hline \hline
Eqs.  & (\ref{xomega0})&(\ref{xnu}) & (\ref{xlambda}) & (\ref{xnu0}) & (\ref{xomega0}),(\ref{xQ}) \\ \hline
\end{tabular}
\end{center}
\caption{Coarse-grained coeficients for the Symmetric hopping model (SHM) and Asymmetric hopping model (AHM), calculated according to equations given in the 4-th row.\label{t1}}
\end{table}
With the coarse-grained coefficients of Table~\ref{t1} we calculate directly the coefficients of regularization and discretization parameter (see Table~\ref{t2}).
\begin{table}[h!]
\begin{center}
\begin{tabular}{| l | c | c | c | c |}
\hline
Model & $A_0$ & $A_1$ & $A_2$ & $\gamma$\\ \hline\hline
SHM   & 0.600 & 0.139 & -0.040 & 0.406 \\ \hline
AHM   & 0.707 & 0.272 & -0.171 & 0.307 \\ \hline 
\end{tabular}
\end{center}
\caption{Regularizing coefficients of the $\varepsilon$-theta expantion and the discretization parameter $\gamma$ calculated by eqs.~(\ref{nu})--(\ref{nu0}) and eq.~(\ref{gamma}), respectively, using the data of Table~\ref{t1}.\label{t2}}
\end{table}

\bibliography{arxiv-Hansmann-Buceta-2013}

\section*{Acknowledgements}
D.H. thanks to CONICET for its support.

\end{document}